# Generalized Reciprocity Relations in Solar Cells with Voltage-Dependent Carrier Collection: Application to p-i-n Junction Devices

Kasidit Toprasertpong[1,*], Amaury Delamarre[1,2], Yoshiaki Nakano[1,2], Jean-François Guillemoles[2,3], and Masakazu Sugiyama[1,2]

[1]School of Engineering, the University of Tokyo, Bunkyo-ku, Tokyo 113-8656, Japan
[2]NextPV LIA CNRS-RCAST, the University of Tokyo, Meguro-ku, Tokyo 153-8904, Japan
[3]CNRS, UMR IPVF, Photovoltaic Institute of Ile-de-France, Palaiseau 91130, France

*Corresponding author, e-mail: toprasertpong@mosfet.t.u-tokyo.ac.jp

## ABSTRACT

Two reciprocity theorems are important for both fundamental understanding of the solar cell operation and applications to device evaluation: 1) the carrier-transport reciprocity connecting the dark-carrier injection with the short-circuit photocarrier collection and 2) the optoelectronic reciprocity connecting the electroluminescence with the photovoltaic quantum efficiency at short circuit. These theorems, however, fail in devices with thick depletion regions such as p-i-n junction solar cells. By properly linearizing the carrier transport equation in such devices, we report that the dark-carrier injection is related to the photocarrier collection efficiency at the operating voltage, not at short circuit as suggested in the original theorem. This leads to the general form of the optoelectronic reciprocity relation connecting the electroluminescence with the voltage-dependent quantum efficiency, providing correct interpretation of the optoelectronic properties of p-i-n junction devices. We also discuss the validity of the well-known relation between the open-circuit voltage and the external luminescence efficiency. The impact of illumination intensity and device parameters on the validity of the reciprocity theorems is quantitatively investigated.

## I. INTRODUCTION

The optoelectronic reciprocity relation [1] has been proposed as a theorem which relates the electroluminescence (EL) and the photovoltaic external quantum efficiency (EQE) at short circuit in p-n junction diodes. With the use of the theorem, several new techniques for evaluating the electrical properties of solar cells based on the optical measurement become available, including the indirect measurement of the subcell voltage of multi-junction solar cells [2-4] and the spatial mapping of the local voltage and other electrical properties [5-10]. The derivation of this reciprocity between the EL (carrier injection followed by photon emission) and the EQE (photon absorption followed by carrier collection) is based on two relations: the reciprocity of photon emission/absorption and the reciprocity of carrier injection/collection. The photonic reciprocity has been originally proved by the detailed balance principle and the ray optic approach [1], and has been recently discussed with more rigorous physics considering the coupling between the photonic and electronic states [11].

On the other hand, the carrier-transport reciprocity, or sometimes called the Donolato theorem, has been first discussed in Ref. [12], describing the symmetry between the injection efficiency of minority carriers under applied voltage and the collection efficiency of photogenerated carriers under short-circuit condition. Despite many further attempts to generalize the carrier-transport reciprocity [13-17], all of them have focused on the carrier dynamics in the quasi-neutral region while neglecting the depletion region, where the carrier dynamics is comparatively complicated. This approach is acceptable in most p-n junction solar cells in which the depletion region is thin compared to other active layers.

However, the exclusion of the depletion region is not an appropriate approach for describing p-i-n junction solar cells, in which the intrinsic region (i-region) is inserted in between the n- and p-regions to extend the depletion region. The p-i-n configuration is usually employed for materials with poor carrier diffusion lengths, such as amorphous and microcrystalline silicon [18,19], complicated alloys [20,21], and quantum structures [22]. The photocarriers are driven by the internal electric field inside the depletion region in addition to the diffusion process, enhancing the carrier collection efficiency and thus the output photocurrent. Since most photogeneration and recombination take place in the depletion region, the carrier-transport reciprocity becomes invalid in p-i-n junction solar cells, and consequently results in the failure of the optoelectronic reciprocity relation. The invalidity of the theorem in p-i-n junction solar cells has been addressed in the original paper [1], and subsequently confirmed by the numerical simulation [23,24]. When evaluating p-i-n junction solar cells using the optical measurement [25], a careful interpretation of the optoelectronic reciprocity relation is needed to extract the correct information of the device electrical properties.

In this paper, we propose the generalized form of the reciprocity relations which is applicable in both p-n and p-i-n junction solar cells and discuss their validity in details. In Sec. II, the original reciprocity relations and the underlying assumptions are briefly discussed. In Sec. III, we show that, even in devices with a thick depletion region, it is possible to linearize the carrier-transport equation by considering well-suited regions, which provides a new aspect of dark-carrier injection and photocarrier collection. In Sec. IV, we derive the generalized form of the optoelectronic reciprocity relation connecting EL and EQE. Readers who are interested in the application on the EL analysis may focus on this section. The requirement of the illumination intensity and the device parameters to keep the reciprocity relations valid is discussed in Sec. V and our findings are summarized in Sec. VI.

## II. ORIGINAL RECIPROCITY RELATIONS

In a p-n junction device, the EL and the photovoltaic EQE at normal emission/incidence, as shown in Figs. 1(a)-(b), are connected by the optoelectronic reciprocity relation [1],

$$\Phi_{\text{EL}}^{(p-n)}(\varepsilon, V) = \text{EQE}^{(\text{sc})}(\varepsilon) \Phi_{\text{bb}}(\varepsilon) \left( e^{qV/kT} - 1 \right), \quad (1)$$

where $\varepsilon$ is the photon energy, $\Phi_{\text{bb}}(\varepsilon)$ is the blackbody radiation flux spectrum, $V$ is the external applied voltage, $q$ is the elementary charge, and $kT$ is the thermal energy. The superscript (sc) is the notation for the short-circuit condition. The optoelectronic reciprocity above has been derived using the carrier-transport reciprocity, which connects the collection efficiency $f_c^{(\text{sc})}$ of carriers photogenerated at a given point $\boldsymbol{r_g}$ under short-circuit condition—the situation with only optical perturbation—with the injection efficiency $f_i$ of excess minority carriers $\delta n_{\text{minor}}$ from the junction $\boldsymbol{r_j}$, i.e. the depletion edge, to the point $\boldsymbol{r_g}$ under applied voltage $V$ without illumination—the situation with only electrical perturbation [12-17],

$$f_c^{(\text{sc})}(\boldsymbol{r_g}) = \left. \frac{\delta n_{\text{minor}}(\boldsymbol{r_g}, V)/n_0(\boldsymbol{r_g})}{\delta n_{\text{minor}}(\boldsymbol{r_j}, V)/n_0(\boldsymbol{r_j})} \right|^{(p-n)}$$
$$= f_i(\boldsymbol{r_g}, V), \quad (2)$$

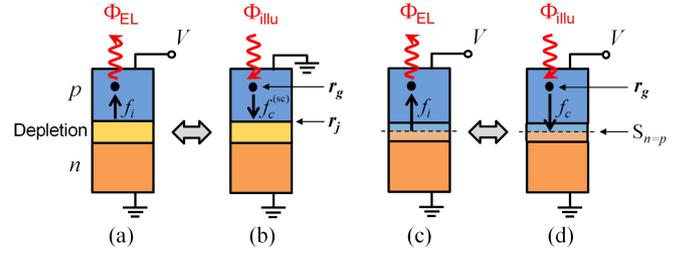

FIG. 1. Reciprocity relation connecting the electroluminescence (carrier injection/photon emission) and photovoltaic EQE (photon absorption/ photocarrier collection). The dynamics of electrons in the p-region is shown in the figure, whereas the hole dynamics in the n-region is in a similar manner. $\boldsymbol{r_g}$ denotes the point of photogeneration, $\boldsymbol{r_j}$ the depletion edge, and $S_{n=p}$ the surface where $n = p$ (see Fig. 2). (a)-(b) Original theorem for p-n junction solar cells considering EQE at short circuit. (c)-(d) Proposed generalized theorem considering EQE at operating voltage.

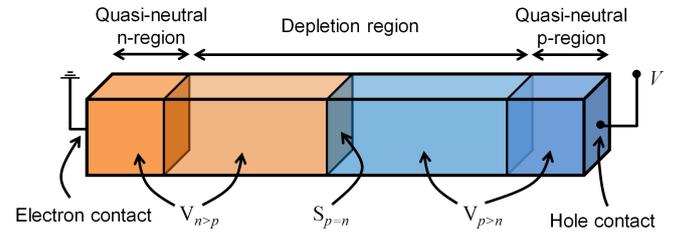

FIG. 2. Device schematic and region notations of a solar cell with a thick depletion region.

where $n_0(\boldsymbol{r_g})$ is the minority carrier density at thermal equilibrium. Although it has been mentioned in some literatures [16,17] that the impact of the drift transport by the electric field is included in the above relation, it does not account for the electric field that depends on the applied voltage, i.e. the electric field in the depletion region, as will be discussed in this study.

## III. GENERALIZED TRANPORT RECIPROCITY

The carrier-transport reciprocity in Eq. (2) has been derived by focusing on only the quasi-neutral region, where the density of one type of carriers, the majority carriers, is practically determined by the doping profile and there is only the dynamics of the minority carriers that has to be considered. On the other hand, an essential difference in the depletion region is that it is necessary to consider both the electron density $n$ and the hole density $p$. Here, we investigate carrier transport dynamics in a device with a thick depletion region shown in Fig. 2, where the depletion region can be either an undoped i-region or a space-charge region of a depleted p- or n-region. To provide a clear picture and quantitative discussion on the theoretical analysis below, which is generalized to devices with arbitrary shapes, a simple p-i-n junction solar cell (Table I) was simulated using the PVcell 1D device simulator [26] as a demonstration (Figs. 3-8).

TABLE I. Material parameters for numerical simulation of the p-i-n junction solar cell in Figs. 3-8

| Parameters | Symbols | Values |
|---|---|---|
| n- and p-region thickness | $l_p, l_n$ | 100 nm |
| i-region thickness | $l_i$ | 500 nm |
| Bandgap | $E_g$ | 1.4 eV |
| Carrier mobility | $\mu_n, \mu_p$ | 0.1 cm$^2$/Vs |
| SRH lifetime | $\tau_n, \tau_p$ | 100 ns |
| Radiative recombination coefficient | $B$ | $10^{-10}$ cm$^3$s$^{-1}$ |
| Auger recombination coefficient | $C_n, C_p$ | $10^{-30}$ cm$^6$s$^{-1}$ |
| Surface recombination velocities | $v_{s,n}, v_{s,p}$ | 0 |
| Intrinsic carrier density | $n_i$ | $10^7$ cm$^{-3}$ |
| Temperature | $T$ | 300 K |

Three types of recombination $R$ are considered: the Shockley-Read-Hall (SRH) recombination, the band-to-band radiative recombination, and the Auger recombination. The SRH recombination rate is given by [27]

$$R_{\text{SRH}} = \frac{np - n_i^2}{\tau_p(n+n_t) + \tau_n(p+p_t)}, \quad (3)$$

where $n_i$ is the intrinsic carrier density, $n_t$ and $p_t$ are the carrier densities when the Fermi level aligns at the trap states, and $\tau_n$ and $\tau_p$ are the SRH lifetimes. The $n$ and $p$ subscripts correspond to the parameters for electrons and holes, respectively, and these notations will be used thereafter. In the depletion region, either doped or non-doped, charge carriers are built up in the steady state and can be divided into the region that $n>p$, i.e. the electron-rich region (V$_{n>p}$ in Fig. 2), and the region that $p>n$, i.e. the hole-rich region (V$_{p>n}$) [28]. (Therefore, strictly speaking, the word "depletion region" is not the correct word use. We keep using this word as it is the standard word for indicating the region between the two quasi-neutral regions.) As can be seen in Fig. 3(c) showing the carrier distribution of a p-i-n junction solar cell described in Fig. 3(a) and Table I, the both carrier densities vary exponentially in the depth direction of the depletion region. The region where the values of $p$ and $n$ are in the same order of magnitude is very thin, allowing the approximation $n \gg p$ and $p \gg n$ in the regions V$_{n>p}$ and V$_{p>n}$, respectively. Furthermore, for deep-level defects, $n_t$ and $p_t$ are small and can be neglected. In this way, $R_{\text{SRH}}$ at the position $\mathbf{r}$ can be approximated by

$$R_{\text{SRH}} \approx \begin{cases} \left(p - \dfrac{n_i^2}{n}\right)\Big/\tau_p & ; \mathbf{r} \in V_{n>p} \\ \left(n - \dfrac{n_i^2}{p}\right)\Big/\tau_n & ; \mathbf{r} \in V_{p>n} \end{cases}, \quad (4)$$

which is confirmed in Fig. 3(d). The radiative recombination is given by

$$R_{\text{rad}} = B(np - n_i^2), \quad (5)$$

where $B$ is the radiative recombination coefficient. Even though the Auger recombination can be large in the quasi-neutral region of some materials, it is generally small and can be omitted in the depletion region where the carrier density is comparatively low. Nevertheless, to generalize the formula which holds in both the quasi-neutral and the depletion regions, we incorporate the Auger recombination process, which can be expressed using the coefficient $C_n$ and $C_p$ [29]

$$R_{\text{Aug}} = C_n n(np - n_i^2) + C_p p(np - n_i^2)$$
$$\approx \begin{cases} C_n n(np - n_i^2) & ; \mathbf{r} \in V_{n>p} \\ C_p p(np - n_i^2) & ; \mathbf{r} \in V_{p>n} \end{cases}. \quad (6)$$

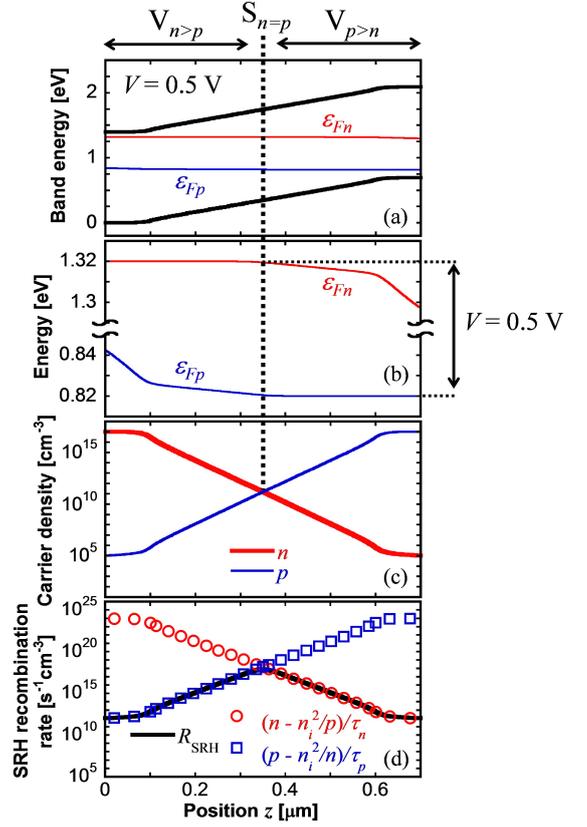

FIG. 3. Simulation of (a) band structure, (b) quasi-Fermi levels, (c) carrier distribution, and (d) SRH recombination rate with the open symbols showing the approximation with Eq. (4). A p-i-n junction solar cell with the parameters summarized in Table I at 0.5 V is simulated as a demonstration.

The approximation in the last line uses that fact that the carrier distribution varies abruptly in the depletion region, similarly to the approximation in Eq. (4).

Electron current density $\mathbf{J_n}$ flowing in the semiconductor can be expressed in terms of drift and diffusion currents,

$$\mathbf{J_n} = q\mu_n n\mathbf{E} + qD_n\nabla n, \quad (7)$$

where $\mu_n$ is the electron mobility, $D_n = \mu_n kT/q$ is the electron diffusion coefficient, and $\boldsymbol{E}$ is the electric field. For materials with non-uniform electron affinity, bandgap, or density of states, the effect of the non-uniformity can be included in $\boldsymbol{E}$ as the effective field [30]. In the absence of illumination, we obtain the carrier transport equation from the current continuity and the recombination in Eqs. (4)-(6):

$$\begin{aligned} 0 &= -\nabla \cdot (q\mu_n n\boldsymbol{E} + qD_n\nabla n) + qR \\ &= -\nabla \cdot (q\mu_n n\boldsymbol{E} + qD_n\nabla n) \\ &\quad + \begin{cases} q\left(p - \dfrac{n_i^2}{n}\right)\left(\dfrac{1}{\tau_p} + Bn + C_n n^2\right) ; \boldsymbol{r} \in \mathrm{V}_{n>p} \\ q\left(n - \dfrac{n_i^2}{p}\right)\left(\dfrac{1}{\tau_n} + Bp + C_p p^2\right) ; \boldsymbol{r} \in \mathrm{V}_{p>n} \end{cases} \end{aligned} \quad (8)$$

Here, $\mu$, $\tau$, $B$, $C$, and $n_i$ can be non-uniform as a function of position $\boldsymbol{r}$.

It is practical to assume that the carrier recombination $R$ has a negligibly small impact on the distribution of the majority carriers, whose density is comparatively large, whereas its impact on the minority-carrier distribution is still important. The validity of this assumption will be discussed in Sec. V B. In this way, as electrons are the majority carriers in the region $\mathrm{V}_{n>p}$, Eq. (8) can be rewritten by

$$0 = -\nabla \cdot (q\mu_n n\boldsymbol{E} + qD_n\nabla n) + \begin{cases} 0 ; \boldsymbol{r} \in \mathrm{V}_{n>p} \\ q\left(n - \dfrac{n_i^2}{p}\right)\left(\dfrac{1}{\tau_n} + Bp + C_p p^2\right) ; \boldsymbol{r} \in \mathrm{V}_{p>n} \end{cases} \quad (9)$$

or

$$0 = -\nabla \cdot (\mu_n n\nabla \varepsilon_{Fn}) + \begin{cases} 0 ; \boldsymbol{r} \in \mathrm{V}_{n>p} \\ q\left(n - \dfrac{n_i^2}{p}\right)\left(\dfrac{1}{\tau_n} + Bp + C_p p^2\right) ; \boldsymbol{r} \in \mathrm{V}_{p>n} \end{cases} \quad (10)$$

by considering that the current can be expressed by the gradient of the quasi-Fermi levels $\varepsilon_{Fn}$ through $\boldsymbol{J_n} = \mu_n n\nabla \varepsilon_{Fn}$ [30]. Similarly, we can write

$$0 = \nabla \cdot (\mu_p p\nabla \varepsilon_{Fp}) + \begin{cases} q\left(p - \dfrac{n_i^2}{n}\right)\left(\dfrac{1}{\tau_p} + Bn + C_n n^2\right) ; \boldsymbol{r} \in \mathrm{V}_{n>p} \\ 0 ; \boldsymbol{r} \in \mathrm{V}_{p>n} \end{cases} \quad (11)$$

for the hole current continuity. The relations $\nabla \varepsilon_{Fn} = 0$ and $\nabla \varepsilon_{Fp} = 0$ are conditions that make the upper line of Eq. (10) and the lower line of Eq. (11) satisfied. That is,

$$\begin{cases} \varepsilon_{Fn}(\boldsymbol{r}) = \text{constant for } \boldsymbol{r} \in \mathrm{V}_{n>p} \\ \varepsilon_{Fp}(\boldsymbol{r}) = \text{constant for } \boldsymbol{r} \in \mathrm{V}_{p>n} \end{cases} . \quad (12)$$

This implies that the majority-carrier quasi-Fermi levels are flat not only in the quasi-neutral region, but also in the depletion region. The constancy of the quasi-Fermi levels of the majority carriers in the depletion region is confirmed by the simulation in Fig. 3(b).

Equation (12) tells us that the quasi-Fermi level splitting $\Delta\varepsilon_F = \varepsilon_{Fn} - \varepsilon_{Fp}$ at the position $\boldsymbol{r_{n=p}}$ in the interface area $\mathrm{S}_{n=p}$ (see Fig. 2) between the regions $\mathrm{V}_{n>p}$ and $\mathrm{V}_{p>n}$, where $n = p$, is given by the applied voltage $V$:

$$\begin{aligned} \Delta\varepsilon_F(\boldsymbol{r_{n=p}}) &\equiv \varepsilon_{Fn}(\boldsymbol{r_{n=p}}) - \varepsilon_{Fp}(\boldsymbol{r_{n=p}}) \\ &= qV \quad \text{for } \boldsymbol{r_{n=p}} \in \mathrm{S}_{n=p} \end{aligned} \quad (13)$$

[See Fig. 3(b)]. By reminding the relation between the carrier densities and the quasi-Fermi level splitting $np = n_i^2 e^{\Delta\varepsilon_F/kT}$, the electron and hole densities at this position are given by

$$n_{n=p} \equiv n(\boldsymbol{r_{n=p}}) = p(\boldsymbol{r_{n=p}}) = n_i e^{qV/2kT}. \quad (14)$$

By introducing the variables

$$u \equiv np/n_i^2 - 1 = e^{\Delta\varepsilon_F/kT} - 1, \quad (15)$$

we can reexpress the electron photocurrent density at $\boldsymbol{r} \in \mathrm{V}_{p>n}$ as

$$\begin{aligned} \boldsymbol{J_n} &= \mu_n n\nabla \varepsilon_{Fn} \\ &= \mu_n n\nabla (\varepsilon_{Fn} - \varepsilon_{Fp}) \\ &= \mu_n n\dfrac{kT\nabla u}{u+1} = \dfrac{\mu_n kT n_i^2 \nabla u}{p}, \end{aligned} \quad (16)$$

{Note that $\nabla\varepsilon_{Fp} = 0$ at $\boldsymbol{r} \in \mathrm{V}_{p>n}$ [Eq. (12)]} and rewrite Eq. (10) for $\boldsymbol{r} \in \mathrm{V}_{p>n}$ as

$$0 = -\nabla \cdot (\dfrac{\mu_n kT n_i^2 \nabla u}{p}) + \dfrac{uqn_i^2}{p}\left(\dfrac{1}{\tau_n} + Bp + C_p p^2\right). \quad (17)$$

Under applied voltage $V$, we define the operator

$$\boldsymbol{L_{n,V}} \equiv -\nabla \cdot (\dfrac{\mu_n kT n_i^2}{p}\nabla) + \dfrac{qn_i^2}{p}\left(\dfrac{1}{\tau_n} + Bp + C_p p^2\right). \quad (18)$$

Using this operator, we can say that in the device at voltage $V$ under dark condition [see Fig. 1(c)], the spatial profile of the parameter $u(\boldsymbol{r})$, denoted by $u_{D,V}(\boldsymbol{r})$, has to satisfy

$$\boldsymbol{L_{n,V}}\left[u_{D,V}(\boldsymbol{r})\right] = 0 \quad \text{for } \boldsymbol{r} \in \mathrm{V}_{p>n}. \quad (19)$$

In the next step, we consider the situation where carriers are additionally photogenerated at point $r_g$ inside $V_{p>n}$ with the delta-function-like generation rate of $G(r) = g\delta(r - r_g)$, while the applied voltage $V$ is kept the same [see Fig. 1(d)]. Provided that the carrier generation is low enough that the majority-carrier density ($p$ in $V_{p>n}$ and $n$ in $V_{n>p}$) is unaffected (the validity is discussed in Sec. V A), $u(r)$ with the photogeneration at the applied voltage $V$, denoted by $u_{L,V}(r, r_g)$, has to satisfy

$$L_{n,V}[u_{L,V}(r, r_g)] = qg\delta(r - r_g) \text{ for } r, r_g \in V_{p>n}. \quad (20)$$

As suggested in Ref. [17], the Green's identity is useful to analyze the boundary of the carrier transport equation. To investigate the Green's identity for the operator $L_{n,V}$, let's consider the integral of $u_{D,V} L_{n,V}[u_{L,V}] - u_{L,V} L_{n,V}[u_{D,V}]$ over the region $V_{p>n}$ (see the Supplemental Material [31] for more details of the derivation):

$$\int_{V_{p>n}} \left( u_{D,V} L_{n,V}[u_{L,V}] - u_{L,V} L_{n,V}[u_{D,V}] \right) dV$$
$$= -\int_{V_{p>n}} \nabla \cdot \left( u_{D,V} \frac{\mu_n kT n_i^2}{p} \nabla u_{L,V} - u_{L,V} \frac{\mu_n kT n_i^2}{p} \nabla u_{D,V} \right) dV$$
$$= -\int_{S_{p=n}} \left( u_{D,V} J_{n,L,V} - u_{L,V} J_{n,D,V} \right) \cdot dS$$
$$= -u_{D,V}(r_{p=n}) \int_{S_{p=n}} \left( J_{n,L,V} - J_{n,D,V} \right) \cdot dS. \quad (21)$$

Here, $J_{n,L,V}$ and $J_{n,D,V}$ are the electron current densities at voltage $V$ with and without illumination, respectively. The last line of Eq. (21) uses the result from Eq. (13) that $\Delta\varepsilon_F$, thus $u$, is constant in the surface $S_{n=p}$ for the given external voltage $V$, provided that both carrier densities $n$ and $p$ at $r_{n=p}$ are not affected by the illumination (see Sec. V A).

The $J_{n,L,V} - J_{n,D,V}$ term corresponds to the increment of current density induced by the illumination at the fixed voltage $V$, or in other words, the *photocurrent* density. By noting that the electron photocurrent density vector $J_{n,L,V} - J_{n,D,V}$ has the opposite direction of the normal vector $dS$ from $V_{p>n}$, thus $(J_{n,L,V} - J_{n,D,V}) \cdot dS < 0$, we can say that the term

$$-\int_{S_{p=n}} \left( J_{n,L,V} - J_{n,D,V} \right) \cdot dS = I_{ph}(r_g, V) \quad (22)$$

is the electron photocurrent that is photogenerated at the point $r_g \in V_{p>n}$ and can be subsequently extracted out of the hole-rich region $V_{p>n}$. By substituting Eqs. (13), (15), (19), (20), and (22) in the first and the last lines of Eq. (21), we obtain the relation

$$qg(e^{\Delta\varepsilon_{F,D}(r_g, V)/kT} - 1) = I_{ph}(r_g, V)(e^{qV/kT} - 1), \quad (23)$$

where $\Delta\varepsilon_{F,D}(r_g, V)$ is the quasi-fermi level splitting at the position $r_g$ under dark condition for the applied voltage $V$. By following the similar derivation in Eqs. (15)-(23) for $r_g \in V_{n>p}$, it is straightforward to show that the same relation holds as well if carriers are photogenerated in $V_{n>p}$ and thus the validity of Eq. (23) can be extended to the photogeneration at any positions $r$ inside the device, including the depletion region.

By rearranging Eq. (23) to

$$\frac{I_{ph}(r_g, V)}{qg} = \frac{e^{\Delta\varepsilon_{F,D}(r_g, V)/kT} - 1}{e^{qV/kT} - 1}, \quad (24)$$

we obtain an interesting relation between the photocurrent of a device with illumination [Fig. 1(c)] on the left side of the equation and the quasi-Fermi level splitting of the same device without illumination [Fig. 1(d)] on the right side. The left side of Eq. (24) is the ratio of collected photocurrent at voltage $V$ to the photogeneration rate at point $r_g$, which can be defined as the *local carrier collection efficiency*

$$f_c(r_g, V) \equiv I_{ph}(r_g, V)/qg \quad (25)$$

On the other hand, the right side of Eq. (24) describes how well the quasi-Fermi level splitting $\Delta\varepsilon_F$ under carrier injection penetrates from $S_{n=p}$, where $\Delta\varepsilon_F = qV$, into the region $V_{p>n}$ or $V_{n>p}$ [see Fig. 3(b)]. Since the right side of Eq. (24) is the generalized form of the original injection efficiency in Eq. (2), which has been defined as the penetration of the excess minority carrier from the depletion region, it can be considered as the generalized *local carrier injection efficiency*

$$f_i(r, V) \equiv \frac{e^{\Delta\varepsilon_{F,D}(r, V)/kT} - 1}{e^{qV/kT} - 1}. \quad (26)$$

In the quasi-neutral region, where the majority-carrier density is considered constant, $e^{\Delta\varepsilon_{F,D}/kT} - 1 = n_{\text{minor}}/n_0 - 1 = \delta n_{\text{minor}}/n_0$ holds and $f_i$ in Eq. (26) gives $f_i$ defined by Eq. (2). In other words, $f_i$ given by Eq. (26) is a more general form which can describe the behavior in all regions.

From Eqs. (24)-(26), we obtain the *generalized carrier-transport reciprocity*

$$f_c(r_g, V) = f_i(r_g, V) \quad (27)$$

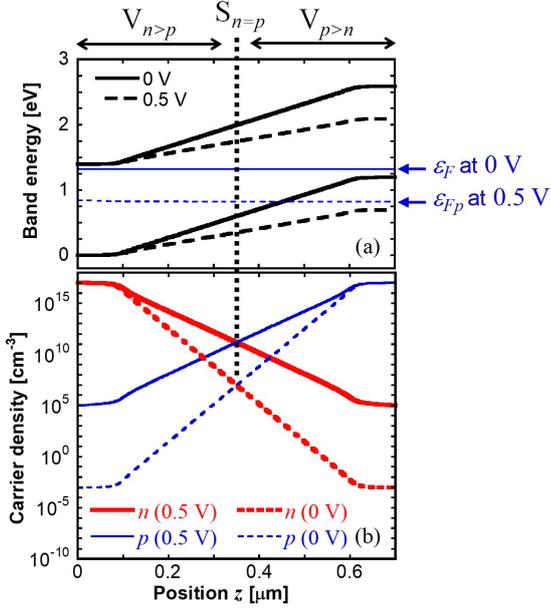

FIG. 4. (a) Band structures and (quasi-)Fermi levels at 0 and 0.5 V. The electron quasi-Fermi level $\varepsilon_{Fn}$ at 0.5 V shown in this scale almost overlaps with the Fermi levels $\varepsilon_F$ at 0 V and is not shown here. (b) Comparison of carrier distribution at 0 V and 0.5 V.

connecting the local dark-carrier injection efficiency with the local photocarrier collection efficiency. Although it has been addressed [1,23,24,32] that, in solar cells with thick depletion regions, the invalidity of the majority/minority-carrier concept in the depletion region may cause the non-linearity in the transport equation and violate the transport reciprocity relation, we have shown that it is possible to divide the depletion region into the electron-rich and hole-rich regions and linearize the carrier transport equation as a linear differential equation of $u$, resulting in the validity of the reciprocity relation.

The significant difference from the original carrier-transport reciprocity is that the photocarrier collection efficiency $f_c$ should not be considered at the short-circuit condition [ $f_c^{(sc)}(r_g)$ in Eq. (2)], but should be defined at the operation voltage $V$ [ $f_c(r_g,V)$ in Eq. (25)]. This is due to the fact that the governing transport equation in Eqs. (10)-(11) at the applied voltage $V \neq 0$ is not the same as that at $V = 0$. Even though Eq. (12) suggests that the quasi-Fermi levels are almost flat in the corresponding majority-carrier regions regardless of the applied voltage [see Fig. 3(b) and Fig. 4(a)], it is shown in Fig. 4(b) that the electron density $n$ in $V_{n>p}$ and the hole density $p$ in $V_{p>n}$ have voltage dependence. Changing the applied voltage $V$ alters the electrostatic potential profile as well as the distance of the bandedges from the quasi-Fermi levels, i.e. $E_C - \varepsilon_{Fn}$ for the conduction band $E_C$ and $E_V - \varepsilon_{Fp}$ for the valence band $E_V$, which in turn alters the majority-carrier density. This results in the inequality

$$L_n^{(sc)} = -\nabla \cdot (\frac{\mu_n kT n_i^2}{p^{(sc)}} \nabla)$$
$$+ \frac{q n_i^2}{p^{(sc)}}\left(\frac{1}{\tau_n} + B p^{(sc)} + C_p \left(p^{(sc)}\right)^2\right)$$
$$\neq L_{n,V} \qquad (28)$$

for $V \neq 0$, which violates the main assumption $L_n^{(sc)} = L_{n,V}$ in the derivation of the original theorem [12-17]. This inequality limits the use of the original theorem [Eq. (2)] on solar cells with thick depletion regions to only small voltage near zero.

The invalidity of the use of the short-circuit carrier collection efficiency can be alternatively explained in terms of the voltage-dependent electric field. At short-circuit condition, the drift-diffusion-based transport equation in Eq. (9) is expressed by

$$0 = -\nabla \cdot (q\mu_n n \mathbf{E}^{(sc)} + qD_n \nabla n) + qR, \qquad (29)$$

where $\mathbf{E}^{(sc)}$ is the electric field profile at $V = 0$. As the electric field in the depletion region decreases with increasing forward bias voltage $V$, carriers follow different carrier transport equations for $V = 0$ and $V \neq 0$, leading to the failure of the simple linear relation between the dynamics of photogenerated carriers at $V = 0$ and electrically injected carriers at $V > 0$.

In addition, the expression for the local dark-carrier injection in the original transport reciprocity relation, the right-hand side of Eq. (2), is not applicable to solar cells containing thick depletion regions since it has been derived based on the assumption $n_{\text{minor}}(r_g,V) = n_0(r_g)e^{\Delta\varepsilon_F/kT}$. This assumption is applicable only if the majority-carrier density remains unchanged from the thermal equilibrium, which holds in the highly-doped quasi-neutral regions but not in the depletion region [Fig. 4(b)]. Instead, the local carrier injection efficiency expressed by the penetration of the quasi-Fermi level splitting, Eq. (26), is the general form which can be used in both the quasi-neutral and the depletion regions. Hence, it is more straightforward to interpret the carrier-transport reciprocity as the relation between the photocarrier collection efficiency $f_c$ and the uniformity of the dark quasi-Fermi level splitting $\Delta\varepsilon_{F,D}$: solar cells with efficient collection of photocarriers under illumination will have the spatially uniform $\Delta\varepsilon_{F,D}$ under the carrier injection mode, and in the opposite way, the inefficient photocarrier collection directly corresponds to the low $\Delta\varepsilon_{F,D}$ under carrier injection.

The carrier-transport reciprocity is confirmed by the 1D simulation of the p-i-n device described in Table I. Fig. 5(a) shows the injection efficiency $f_i$ to the n-region at the depth $z_g$ = 30 nm and the collection efficiency $f_c$ of carriers photogenerated in a thin layer at the same depth, simulated using Eq. (25) and Eq. (26), respectively. Fig. 5(b) is similar but is simulated at the depth $z_g$ = 500 nm, which is inside the i-region. It can be seen that the voltage-dependent collection efficiency $f_c$, not the collection efficiency $f_c^{(sc)}$ at short circuit, agrees well with the value of dark injection, confirming the proposed transport reciprocity relation in Eq. (27).

## IV. GENERALIZED OPTOELECTRONIC RECIPROCITY

When the device at applied voltage $V$ is illuminated with the monochromatic light with photon energy $\varepsilon$, by using the probability $a(\mathbf{r},\varepsilon,V)d\mathbf{r}$ that the incident photon is absorbed in the volume element $d\mathbf{r}$ at $\mathbf{r}$, EQE can be obtained by integrating the photocurrent induced from each point over the entire volume $V_{device}$,

$$\mathrm{EQE}(\varepsilon,V) = \int_{V_{device}} a(\mathbf{r},\varepsilon,V) f_c(\mathbf{r},V) d\mathbf{r}. \tag{30}$$

On the other hand, in the absence of illumination, the net photon flux $\phi_z(\mathbf{r},\varepsilon,V)d\mathbf{r}$ of photon energy $\varepsilon$ that is emitted from the volume element $d\mathbf{r}$ of the absorber and is able to escape from the device surface without being reabsorbed can be expressed using the generalized Kirchhoff's law [33], which has been further generalized for materials with the field-dependent absorption [11],

$$\phi_z(\mathbf{r},\varepsilon,V)d\mathbf{r} = a(\mathbf{r},\varepsilon,V)\Phi_{bb}(\varepsilon)\left(e^{\Delta\varepsilon_{F,D}(\mathbf{r},V)/kT}-1\right)d\mathbf{r}. \tag{31}$$

The total electroluminescence $\Phi_{EL}$ from the absorber surface can be obtained by integrating the photon flux $\phi_z(\mathbf{r},\varepsilon,V)d\mathbf{r}$ contributed from each volume element,

$$\begin{aligned}\Phi_{EL}(\varepsilon,V) &= \int_{V_{device}} a(\mathbf{r},\varepsilon,V)\Phi_{bb}(\varepsilon)\left(e^{\Delta\varepsilon_{F,D}(\mathbf{r},V)/kT}-1\right)d\mathbf{r} \\ &= \int_{V_{device}} a(\mathbf{r},\varepsilon,V) f_c(\mathbf{r},V)\Phi_{bb}(\varepsilon)\left(e^{qV/kT}-1\right)d\mathbf{r} \\ &= \mathrm{EQE}(\varepsilon,V)\Phi_{bb}(\varepsilon)\left(e^{qV/kT}-1\right),\end{aligned} \tag{32}$$

where the generalized transport reciprocity in Eq. (27) is used to obtain the second line and EQE in Eq. (30) to obtain the last line. Equation (32) represents the general form of the optoelectronic reciprocity which links the EL with the photovoltaic EQE [see Figs. 1(c)-(d)]. By following the similar derivation in Ref. [1], the optoelectronic reciprocity relation in Eq. (32) can be extended to the electroluminescence $\Phi_{EL}^{(tilt)}(\varepsilon,V,\theta)$ emitted at the tilt angle $\theta$, with $\mathrm{EQE}(\varepsilon,V)$ substituted with $\mathrm{EQE}^{(tilt)}(\varepsilon,V,\theta)$

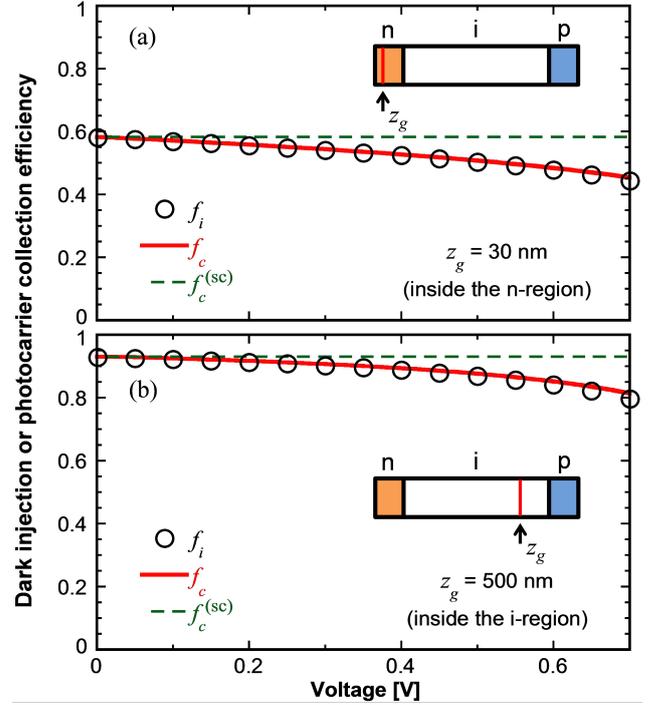

FIG. 5. Simulated local dark-carrier injection and photocarrier collection efficiencies of the p-i-n junction solar cell in (a) the n-region at the depth $z_g$ = 30 nm and (b) the i-region at $z_g$ = 500 nm. Circle symbols show the injection efficiency obtained from the simulated quasi-Fermi level of the device without illumination and Eq. (26). Lines represent the voltage-dependent collection efficiency $f_c$ of carriers photogenerated in a thin layer at the corresponding depth $z_g$ with the photogeneration rate of $10^{10}$ cm$^{-2}$s$^{-1}$, obtained from the simulated photocurrent and Eq. (25). $f_c^{(sc)}$ corresponds to $f_c$ at short circuit. Simplified device schematics indicate the positions of photogeneration.

measured at the light incident angle $\theta$. We would like to emphasize, as it might be easily misinterpreted due to the term $(e^{qV/kT}-1)$, that there is no assumption for the formula of EL that the quasi-Fermi level splitting $\Delta\varepsilon_{F,D}$ must equal to $qV$ in the entire device. The non-uniformity of the quasi-Fermi level splitting $\Delta\varepsilon_{F,D}$ has been included in the $\mathrm{EQE}(\varepsilon,V)$ term.

Equation (32) derived above has the similar expression to the original optoelectronic reciprocity relation in Eq. (1), but the voltage-dependent EQE should be used instead of the short-circuit EQE. This is a consequence of the voltage-dependent carrier collection in the carrier-transport reciprocity [Eq. (27)] and the voltage-dependent light absorptance [11]. Previously, it has been addressed in Ref. [34] that in non-linear solar cells, the differential form of the electroluminescence follows the reciprocity relation $d\Phi_{EL}/d(qV/kT) = \mathrm{EQE}\,\Phi_{bb}e^{qV/kT}$ with EQE represented the voltage-dependent EQE. We have shown in this paper that the carrier transport in the depletion region is linear at fixed voltage and the absolute electroluminescence $\Phi_{EL}$ itself also follows the optoelectronic reciprocity relation with the voltage-dependent EQE.

This finding suggests that when applying the reciprocity relation to analyze the electrical properties of solar cells from optical measurement, for instance extracting the subcell voltage of multi-junction solar cells and mapping the voltage distribution from emitted EL [2-10], the voltage-dependent EQE data are required for correct data analysis. The original reciprocity relation expressed by Eq. (1) may be used only in solar cells with EQE independent of voltage such as p-n junction solar cells with sufficiently high doping concentration.

## V. VALIDITY DISCUSSION

### A. Limitation of illumination intensity

One main assumption in the derivation above is the low illumination intensity such that the majority-carrier density remains unchanged. It is important to quantitatively estimate such limit as it is directly related to the limitation of the illumination intensity in the EQE measurement. If the electric field in the depletion region is sufficiently high, the transport of photogenerated carriers is dominated by the drift transport. For the photogeneration in $V_{p>n}$, the electron photocurrent density $J_{ph,n}$ and the electron density increment $\Delta n$ due to illumination are related by [28]

$$J_{ph,n} = q\mu_n \Delta n E. \qquad (33)$$

The similar relation of the hole photocurrent density $J_{ph,p}$ and the hole density increment $\Delta p$ for the photogeneration in $V_{n>p}$ can also be obtained.

Remind that the electron and hole densities at $r_{n=p} \in S_{n=p}$ given by $n_{n=p} = n_i e^{qV/2kT}$ [Eq. (14)] are the minimum values of the majority-carrier density in both the regions $V_{p>n}$ and $V_{n>p}$. The condition for keeping the majority-carrier distribution unaffected by the induced photocarriers $\Delta n$ is estimated by

$$J_{ph,n\,\text{or}\,p}^{\perp}(r_{n=p}) < q\mu_{n\,\text{or}\,p} n_i e^{qV/2kT} E_{n=p}^{\perp}, \qquad (34)$$

where $J_{ph,n\,\text{or}\,p}^{\perp}(r_{n=p})$ is the electron or hole photocurrent density normal to the surface $S_{n=p}$ and $E_{n=p}^{\perp}$ is the electric field normal to $S_{n=p}$. It can be interpreted from Eq. (34) that in materials where $\mu_n > \mu_p$, the theorem validity is more tolerant to photogeneration in $V_{p<n}$ than photogeneration in $V_{n>p}$ as electron photocurrent density $J_{ph,n}^{\perp}$ induces smaller steady-state charge density $\Delta n$.

For planar devices, the output photocurrent density $J_{ph}$ is given by $J_{ph} = J_{ph,n}^{\perp}(r_{n=p}) + J_{ph,p}^{\perp}(r_{n=p})$ and the electric field $E_{n=p}$ is normal to $S_{n=p}$. Then,

$$J_{ph} < qn_i E_{n=p} e^{qV/2kT} \min(\mu_n, \mu_p) \qquad (35)$$

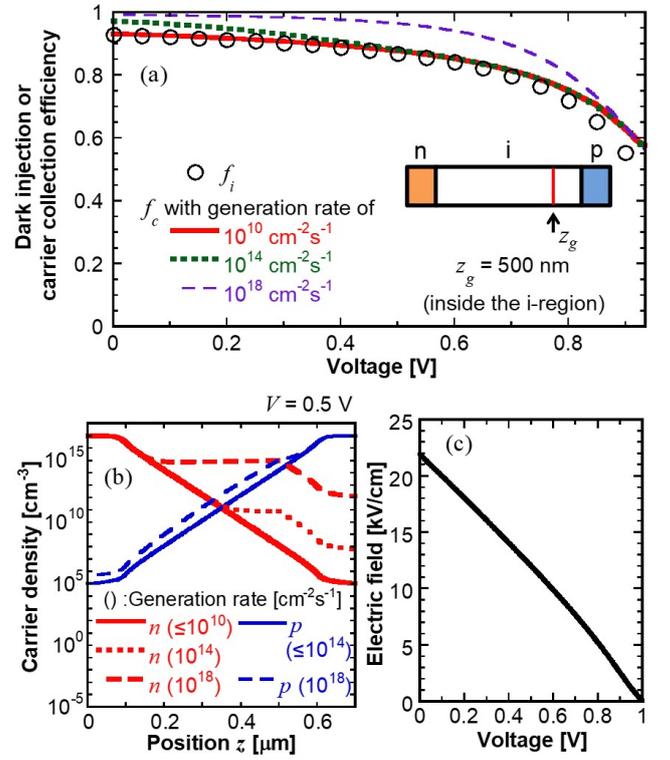

FIG. 6. (a) Failure of the carrier-transport reciprocity at high illumination intensity and high applied voltage for the photogeneration at $z_g = 500$ nm. (b) Electron density $n$ and hole density $p$ at 0.5 V under different photogeneration rates. The distributions of $n$ at the photogeneration rate of $10^{10}$ cm$^{-2}$s$^{-1}$, and $p$ at $10^{10}$ and $10^{14}$ cm$^{-2}$s$^{-1}$ mostly overlap with the distributions under dark condition. The photogeneration rate as high as $10^{18}$ cm$^{-2}$s$^{-1}$, roughly equivalent to 5-sun intensity, affects $n$ in the electron-rich region and $p$ in the hole-rich region. (c) Electric field $E_{n=p}$ at $z_{n=p}$.

is a sufficient condition to make Eq. (34) simultaneously holds for both the electron and hole photocurrents, ensuring the validity of the transport reciprocity. In the EQE measurement, this is equivalent to the condition for the illuminated photon flux

$$\Phi_{illu,EQE}(\varepsilon) < \frac{n_i E_{n=p} e^{qV/2kT} \min(\mu_n, \mu_p)}{\text{EQE}(\varepsilon, V)}. \qquad (36)$$

For the p-i-n junction solar cell in Table I, Eq. (35) gives the photocurrent limit at $V = 0$ V [ $E_{n=p} = 22$ kV/cm as shown in Fig. 6(c)] to be 3.5 nA/cm$^2$ and at $V = 0.5$ V ( $E_{n=p} = 12$ kV/cm ) to be 30 μA/cm$^2$, which are equivalent to the carrier fluxes of $2\times 10^{10}$ cm$^{-2}$s$^{-1}$ and $2\times 10^{14}$ cm$^{-2}$s$^{-1}$, respectively. These values well agree with Fig. 6(a) showing that the photogeneration rate of $1\times 10^{14}$ cm$^{-2}$s$^{-1}$ is too high for the validity of the reciprocity relation at 0 V, and $1\times 10^{18}$ cm$^{-2}$s$^{-1}$ is too high for both 0 V and 0.5 V. The carrier distribution under illumination at 0.5 V shown in Fig. 6(b) illustrates that this is due to the remarkable impact of high photogeneration rate on the majority-carrier distribution: the generation rate of $10^{18}$ cm$^{-2}$s$^{-1}$, but not $10^{14}$ cm$^{-2}$s$^{-1}$, affects the distribution of the majority-carrier density at 0.5 V. See also the Supplemental Material [31] for the behavior of the carrier distributions as well as the quasi-Fermi levels under other device conditions.

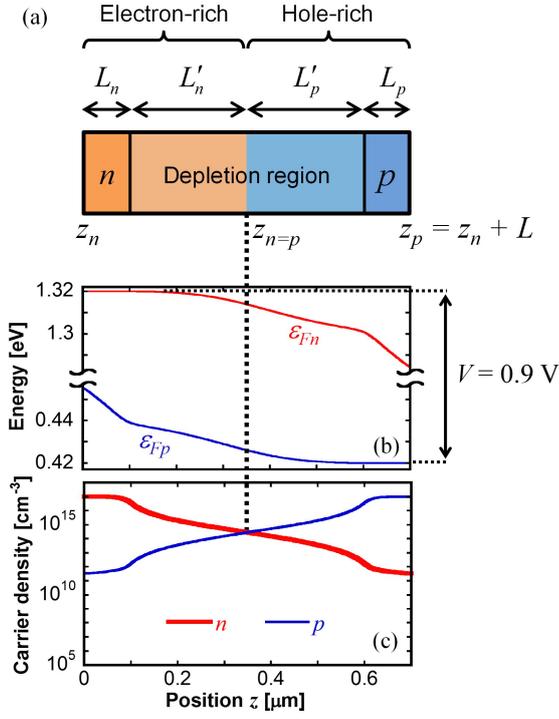

FIG. 7. (a) Position and width notations for the one-dimensional device discussed in Sec. V B. (b) Quasi-Fermi levels and (c) carrier distribution in the p-i-n device described by Table I at applied voltage of 0.9 V.

## B. Requirement of electric field and mobility

Another assumption for the validity of the reciprocity relations is that the quasi-Fermi level of the majority carriers should be flat under carrier injection [Eq. (12)]. Here, we investigate the spatial variation of the electron quasi-Fermi level $\varepsilon_{Fn}$ in the electron-rich region by including the recombination term in the upper line of Eq. (10). To investigate such impact in a quantitative manner, we simplify the discussion to a one-dimensional device with uniform material properties of $\mu_n$, $\tau_n$, $\tau_p$, $B$, $C_n$, $C_p$, and $n_i$, and no surface recombination. The 1D transport equation for electrons in the electron-rich region ($z_n \le z < z_{n=p}$ in Fig. 7, where $z_n$ is the position of the electron contact and $z_{n=p}$ is the position where $n = p$) gives

$$\frac{d}{dz}\left(\mu_n n \frac{d}{dz}\varepsilon_{Fn}\right) = qR(z)$$
$$= q\left(p - \frac{n_i^2}{n}\right)\left(\frac{1}{\tau_p} + Bn + C_n n^2\right). \quad (37)$$

As Eq. (37) cannot be easily solved analytically, we employ some approximations to estimate a rough tendency of $\varepsilon_{Fn}$. For the first-order approximation, we first express $n$ and $p$ by $n^{(0)}$ and $p^{(0)}$ assuming the flat quasi-Fermi levels $\varepsilon_{Fn}^{(0)}$, and then solve for the first-order solution $\varepsilon_{Fn}^{(1)}$. Even though this approach cannot give the exact solution of $\varepsilon_{Fn}(z)$, it provides a rough estimation of how the device parameters, such as mobilities, lifetimes, and electric field, affect the uniformity of the majority-carrier quasi-Fermi levels.

For non-degenerate carriers, the spatial distribution of $n^{(0)}$ follows $\ln n^{(0)} \propto -(E_C - \varepsilon_{Fn}^{(0)})/kT \propto -E_C/kT$ for the flat quasi-Fermi level. For the electric field $E_{n=p}$ locally uniform near $z_{n=p}$, it is a fair estimation to say

$$n^{(0)}(z) \approx n_{n=p} e^{-qE_{n=p}(z-z_{n=p})/kT} \quad (38)$$

in the vicinity of $z_{n=p}$. By using Eq. (14) and $n^{(0)}p^{(0)} - n_i^2 = n_i^2(e^{qV/kT} - 1) \approx n_i^2 e^{qV/kT}$ for large $V$, we can integrate Eq. (37) from $z_n$ to $z$ near $z_{n=p}$ and obtain

$$\left[\mu_n n \frac{d}{dz'}\varepsilon_{Fn}(z')\right]_{z_n}^{z} = q\int_{z_n}^{z} R(z')dz'$$

$$\mu_n n^{(0)} \frac{d}{dz}\varepsilon_{Fn}^{(1)}(z) - J_{n0} = \int_{z_n}^{z} \frac{q}{\tau_p n^{(0)}}\left(n^{(0)}p^{(0)} - n_i^2\right)dz'$$
$$+ \int_{z_n}^{z} qB\left(n^{(0)}p^{(0)} - n_i^2\right)dz'$$
$$+ \int_{z_n}^{z} qC_n n^{(0)}\left(n^{(0)}p^{(0)} - n_i^2\right)dz'$$
$$\approx \frac{n_i kT}{\tau_p E_{n=p}} e^{qV/2kT} e^{qE_{n=p}(z-z_{n=p})/kT}$$
$$+ qBn_i^2(z - z_n)e^{qV/kT}$$
$$+ qC_n n_i^2 N_D L_n e^{qV/kT}, \quad (39)$$

where $J_{n0}$ is the electron current density at the electron contact $z_n$ and $L_n$ is the width of the quasi-neutral n-region [see Fig. 7(a)]. The expression for $n^{(0)}$ is approximated by Eq. (38) for the first integral (SRH recombination) as the integrand is large only near $z_{n=p}$, and approximated by the electron density $N_D$ at the undepleted n-region for the third integral (Auger recombination) due to the considerably large $n$ in the quasi-neutral region. In a device without surface recombination, the electron current must vanish at the hole contact $z_p$ and $J_{n0}$ is given by to the total recombination throughout the device:

$$J_{n0} = -q\int_{z_n}^{z_p} R(z)dz$$
$$= -q\int_{z_n}^{z_{n=p}} R(z)dz - q\int_{z_{n=p}}^{z_p} R(z)dz$$
$$\approx -\frac{n_i kT}{E_{n=p}}\left(\frac{1}{\tau_p} + \frac{1}{\tau_n}\right)e^{qV/2kT} - qBn_i^2 L e^{qV/kT}$$
$$- qn_i^2\left(C_n N_D L_n + C_p N_A L_p\right)e^{qV/kT}, \quad (40)$$

where $N_A$ and $L_p$ are the hole density and the width of the quasi-neutral p-region, respectively, and $L$ is the device total thickness. Note that $J_{n0}$ is not a function of position $z$. Then, integrating $d\varepsilon_{Fn}^{(1)}/dz$ using Eqs. (39)-(40) gives the deviation of $\varepsilon_{Fn}^{(1)}$

$$\delta\varepsilon_{Fn}^{(1)} \equiv \varepsilon_{Fn}^{(1)}(z_{n=p}) - \varepsilon_{Fn}(z_n)$$

$$= \int_{z_n}^{z_{n=p}} \frac{d}{dz}\varepsilon_{Fn}^{(1)} dz$$

$$= \int_{z_n}^{z_{n=p}} \frac{1}{\mu_n n^{(0)}}\left[\left(\mu_n n^{(0)}\frac{d}{dz}\varepsilon_{Fn}^{(1)} - J_{n0}\right) + J_{n0}\right]dz$$

$$\approx -\frac{kT}{\mu_n E_{n=p}}\left[\frac{kT}{qE_{n=p}}\left(\frac{1}{\tau_n} + \frac{1}{2\tau_p}\right)\right.$$
$$\left. + Bn_i e^{\frac{qV}{2kT}}\left(L_p + L'_p\right) + n_i C_p N_A L_p e^{\frac{qV}{2kT}}\right], \quad (41)$$

where $L'_p$ is width of the depletion region that has $p > n$. Due to the $1/n$ term, the integrant is particularly large for $z$ near $z_{n=p}$ and the expressions in Eqs. (38) and (39) are used. The deviation $\delta\varepsilon_{Fp}^{(1)}$ for the hole quasi-Fermi level in the hole-rich region can be expressed in the same way by interchanging the subscripts $n$ and $p$.

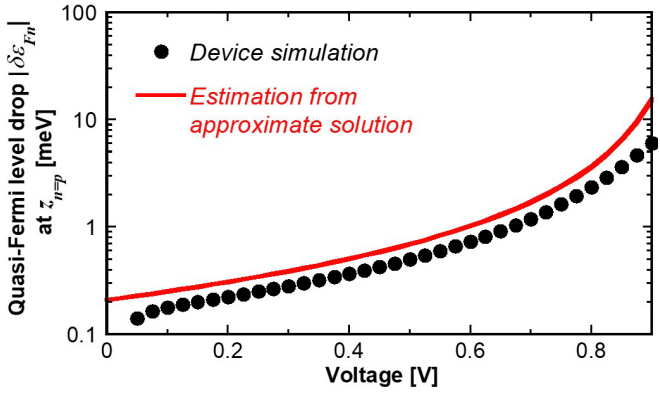

FIG. 8. Electron quasi-Fermi level drop $\delta\varepsilon_{Fn}$ at $z_{n=p}$ under carrier injection mode estimated from Eq. (41) as compared to $\delta\varepsilon_{Fn}$ numerically calculated from the device simulation.

Figure 8 demonstrates $\delta\varepsilon_{Fn}$ at $z_{n=p}$ for the device parameters in Table I under carrier injection mode, for that approximated from Eq. (41) and the exact solution solved by the device simulator employed so far. Even though Eq. (41) results in finite approximation error, which is due to the assumptions made during the derivation, the estimation of $\delta\varepsilon_{Fn}$ by Eq. (41) provided a good approximation within a factor of 2 and can be used as a convenient tool for further understanding of the behavior of $\delta\varepsilon_{Fn}$.

Since the constant quasi-Fermi levels of the majority carriers [Eq. (12)] is the requisite condition for the carrier-transport and the optoelectronic reciprocity relations, both

$\delta\varepsilon_{Fn}$ and $\delta\varepsilon_{Fp}$ should be sufficiently small compared to $kT$ to keep the relations valid, that is,

$$\begin{cases} \frac{1}{\mu_n E_{n=p}}\left[\frac{kT}{qE_{n=p}}\left(\frac{1}{\tau_n} + \frac{1}{2\tau_p}\right)\right. \\ \left. + Bn_i e^{\frac{qV}{2kT}}(L_p + L_{i,p}) + n_i C_p N_A L_p e^{\frac{qV}{2kT}}\right] < 1 \\ \frac{1}{\mu_p E_{n=p}}\left[\frac{kT}{qE_{n=p}}\left(\frac{1}{2\tau_n} + \frac{1}{\tau_p}\right)\right. \\ \left. + Bn_i e^{\frac{qV}{2kT}}(L_n + L_{i,n}) + n_i C_n N_D L_n e^{\frac{qV}{2kT}}\right] < 1 \end{cases}. \quad (42)$$

Particularly in a symmetric device (all parameters have the same values for the subscripts $n$ and $p$), the estimated requirement for the electric field when the SRH process is dominant is

$$E_{n=p} > \sqrt{3kT/2q\mu\tau} \quad (43)$$

and when the radiative or Auger process is dominant is

$$E_{n=p} > \left(\tfrac{1}{2}BL + CN_D L_n\right)n_i e^{qV/2kT}/\mu. \quad (44)$$

The device described by Table I is dominated by the SRH process and Eq. (43) gives the required electric field of 2 kV/cm, which is consistent with the large deviation of the injection efficiency $f_i$ from the collection efficiency $f_c$ in Fig. 6(a) when the electric field $E_{n=p}$ [Fig. 6(c)] approaches 2 kV/cm at 0.9 V. This is confirmed by a significant drop of the quasi-Fermi levels of majority carriers at $z_{n=p}$ for $V = 0.9$ V shown in Fig. 7(b) and Fig. 8. Equations (42)-(44) can also be interpreted as the lower limit of the carrier mobility $\mu$ in order to keep the reciprocity relations valid for given device structure and operating voltage. For instance, the same device requires a mobility at least $3kT/2q\tau(E_{n=p}^\perp)^2 = 160$ cm$^2$/Vs to validate the relations at 1 V, at which the electric field $E_{n=p}^\perp$ is as low as 51 V/cm. This finding is in a good agreement with the simulation results in Ref. [32] suggesting that the reciprocity relations become invalid at high applied voltage, where the electric field becomes weak, and become invalid in the entire voltage range in devices with extremely low carrier mobility.

### C. SRH via shallow defects

The SRH recombination rate in Eq. (4) assumes the deep-level defects. For shallow-level defects, for instance near the conduction band, $n_t$ becomes large and the recombination rate in Eq. (3) becomes

$$R_{\text{SRH}} = \left(np - n_i^2\right)/\tau_p n_t. \qquad (45)$$

This has the same expression as the radiative recombination in Eq. (5) and can similarly be included in the above derivation, provided that the illumination is weak enough that shallow-level defects are not changed to the deep-level type upon illumination.

## D. Application to various types of solar cells

Since the derivation above includes the field-dependent carrier transport in the depletion region, the proposed reciprocity relations are applicable to p-i-n junction solar cells with the i-region having field either uniform or non-uniform, or having field-dependent material parameters such as mobility, lifetime, and density of states. They can also be used to describe the behavior of p-i-n quantum structure solar cells, provided that the carrier mobility in such structures can be defined [35]. Counting for the voltage dependency of EQE also extends the validity of the optoelectronic reciprocity to absorbers with voltage-dependent absorptance [36,37], which is not valid in the original theorem. Moreover in p-n junction solar cells, particularly in low-doped devices, it is well recognized that applying forward bias decreases the depletion width and thus widens the quasi-neutral region. The variation of the depletion width, which has not been taken into account so far, appears as the voltage dependence of carrier collection as well as EQE and can be treated by our modified formulae.

## E. Open-circuit voltage

It is worth noting about the relation of the open-circuit voltage $V_{oc}$ with the external luminescence efficiency $\eta_{\text{ext}}$, which has been discussed in p-n junction solar cells by assuming the superposition of the short-circuit current and the dark diode current and is expressed by [1,23,38]

$$V_{oc}^{(p-n)} = V_{oc}^{(\text{rad})} + \frac{kT}{q}\ln\eta_{\text{ext}}, \qquad (46)$$

where $V_{oc}^{(\text{rad})}$ is the ideal open-circuit voltage in the radiative limit. For devices with voltage-dependent carrier collection such as p-i-n junction solar cells, the non-constant photocurrent at forward bias results in the failure of the simple superposition between the short-circuit and dark currents. Instead, $V_{oc}$ of such devices should be expressed by [32]

$$V_{oc} = V_{oc}^{(\text{rad})} + \frac{kT}{q}\ln\eta_{\text{ext}} + \frac{kT}{q}\ln\frac{F_c(V_{oc})}{F_i(V_{oc})}, \qquad (47)$$

where $F_c(V)$ and $F_i(V)$ are the spatial averages of the local collection efficiency $f_c(\boldsymbol{r},V)$ and injection efficiency $f_i(\boldsymbol{r},V)$, respectively. For the illumination intensity and device parameters that satisfy Eqs. (35) and (42), the carrier-transport reciprocity, $f_c(\boldsymbol{r},V) = f_i(\boldsymbol{r},V)$ and thus $F_c(V) = F_i(V)$, holds. Then we obtain

$$V_{oc} = V_{oc}^{(\text{rad})} + \frac{kT}{q}\ln\eta_{\text{ext}}, \qquad (48)$$

which is the same expression for p-n junction solar cells [Eq. (46)]. The generalized carrier-transport reciprocity results in the open-circuit voltage $V_{oc}$ unaffected by the characteristics of voltage-dependent carrier collection, which is in a good agreement with the simulation result in Ref. [32] reporting the validity of Eq. (48) in depleted organic solar cells having voltage-dependent photocurrent, and with the experimental result in Ref. [39] showing the satisfactorily high $V_{oc}$ in p-i-n quantum well solar cells in spite of their poor photocarrier collection under forward bias.

## VI. CONCLUSION

In this study, we have investigated the validity of the reciprocity relations in solar cells with thick depletion regions. The voltage dependence of the photocarrier collection efficiency and the EQE in such devices results in the failure of the two reciprocity theorems, the carrier-transport reciprocity relating the dark-carrier injection and short-circuit photocarrier collection, and the optoelectronic reciprocity, which is the consequence of the previous relation, relating the EL with the short-circuit EQE. The analysis of carrier dynamics in the depletion region suggests that the original transport reciprocity theorem should be modified: the photocarrier collection efficiency should be defined at the operating voltage, not at short circuit. As a consequence, we obtain the general form of the optoelectronic reciprocity which connects the EL with the EQE at the operating voltage, extending its application, e.g. the EL diagnosis in solar cells, to solar cells with voltage-dependent EQE. This finding extends the validity of another well-known relation between the open-circuit voltage and the external luminescence efficiency to devices with voltage-dependent photocurrent. The acceptable ranges of the illumination intensity, the depletion-region field, and the carrier mobility to assure the validity of the theorems have been discussed and are analytically expressed in terms of the material parameters.

## ACKNOWLEDGMENTS

A part of this study was supported by the Research and Development of Ultra-high Efficiency and Low-cost III-V Compound Semiconductor Solar Cell Modules project under the New Energy and Industrial Technology Development Organization (NEDO). The authors would like to thank Daniel Suchet for the helpful discussions.